# Room-temperature nonlinear transport and microwave rectification in antiferromagnetic MnBi$_2$Te$_4$ films

Shanshan Liu[1#], Rhonald Burgos[2#], Enze Zhang[3], Naizhou Wang[4], Xiao-Bin Qiang[5], Chuanzhao Li,[6] Qihan Zhang[7], Z. Z. Du[5], Rui Zheng[1], Jingsheng Chen[7], Qing-Hua Xu[1], Kai Leng,[6] Weibo Gao[4], Faxian Xiu[3], Dimitrie Culcer[2*], Kian Ping Loh[1*]

[1]Department of Chemistry, National University of Singapore, Singapore 117543, Singapore
[2]School of Physics, The University of New South Wales, Sydney 2052, Australia
[3]State Key Laboratory of Surface Physics and Department of Physics, Fudan University, Shanghai 200433, China
[4]Division of Physics and Applied Physics, School of Physical and Mathematical Sciences, Nanyang Technological University, Singapore 637371, Singapore
[5]Shenzhen Institute for Quantum Science and Engineering and Department of Physics, Southern University of Science and Technology (SUSTech), Shenzhen 518055, China.
[6]Department of Applied Physics, The Hong Kong Polytechnic University, Kowloon 999077 Hong Kong, China
[7]Department of Materials Science and Engineering, National University of Singapore, Singapore, 117575, Singapore

[#] These authors contributed equally to this work.
[*]Correspondence and requests for materials should be addressed to K.P.L. (Email: chmlohkp@nus.edu.sg) and D.C. (Email: d.culcer@unsw.edu.au).

**Abstract**

The discovery of the nonlinear Hall effect provides an avenue for studying the interplay among symmetry, topology, and phase transitions, with potential applications in signal doubling and high-frequency rectification. However, practical applications require devices fabricated on large area thin film as well as room-temperature operation. Here, we demonstrate robust room-temperature nonlinear transverse response and microwave rectification in MnBi$_2$Te$_4$ films grown by molecular beam epitaxy. We observe multiple sign-reversals in the nonlinear response by tuning the chemical potential. Through theoretical analysis, we identify skew scattering and side jump, arising from extrinsic spin-orbit scattering, as the main mechanisms underlying the observed nonlinear signals. Furthermore, we demonstrate radio frequency (RF) rectification in the range of 1-8 gigahertz at 300 K. These findings not only enhance our understanding of the relationship between nonlinear response and magnetism, but also expand the potential applications as energy harvesters and detectors in high-frequency scenarios.

**Introduction**

The Hall effect, which refers to the generation of a transverse voltage in response to longitudinal currents, has led to discoveries in fractional charge and statistics,[1,2] topological phases of materials,[3,4] and potential applications in quantum computing devices.[5] In contrast to the linear Hall effect, where the transverse Hall voltage is linearly dependent on the longitudinal driving current, the recently identified second-order nonlinear Hall effect (NLHE) is characterized by a transverse Hall voltage that varies quadratically with the longitudinal a.c. driving currents.[6] Moreover, NLHE occurs in systems with time-reversal symmetry,[7] unlike the linear Hall effects that require time-reversal symmetry breaking. Various low-symmetry materials, such as atomically thin $WTe_2$[8,9], corrugated bilayer graphene[10], and twisted bilayer $WSe_2$[11], exhibit Berry curvature dipole (BCD)-induced NLHE. However, BCD-induced NLHE requires stringent symmetry conditions such as the breaking of inversion and $C_3$ rotation symmetry,[6,12,13] and its magnitude diminishes sharply as temperature increases due to its sensitivity to electronic band structures.[8,9] As a result, there are only a few reports of room-temperature stable NLHE to date, as exemplified by semimetals $TaIrTe_4$[14] and $BaMnSb_2$[15], semiconductor $BiTeBr$[12], and metal Pt[16]. BCD-induced NLHE is predominantly a low-temperature phenomenon. Therefore, it is highly desirable to explore non-BCD-dependent mechanisms as well as materials that exhibit room-temperature nonlinear second-order effect to enable practical applications.

$MnBi_2Te_4$, a magnetic topological material,[17–21] being the playground for studying quantum anomalous Hall effect and axion insulator states in odd/even layered devices, and nonlinear transport originating from the quantum metric dipole (QMD) has been discovered in ideal even-layer $MnBi_2Te_4$ devices.[22,23] However, the QMD-induced nonlinear Hall effect disappears above the Néel temperature (~23 K). In contrast, antiferromagnet CuMnAs, which possesses *PT*-symmetry, exhibits an intrinsic NLHE with a magnitude several orders larger, which persists above room temperature.[24]

Although disorder effect has often been neglected in the interpretation of NLHE in *PT*-symmetric systems, recent theoretical predictions suggest that disorder can overshadow the intrinsic contribution. [25] In view of this, we investigate the effect of disorder on nonlinear transport by studying molecular beam epitaxy (MBE)-grown $MnBi_2Te_4$ films. Unlike the reported compensated antiferromagnetic (AFM) order in ideal even-layer $MnBi_2Te_4$ flakes, MBE-grown $MnBi_2Te_4$ film exhibits a net magnetization, originating from disorders.[26]

In this study, we demonstrate the observation of nonlinear transverse response in MBE-grown $MnBi_2Te_4$ films that persists to 320 K. Through scaling analysis of the nonlinear transverse magnitude and longitudinal conductivity, we attribute the dominant microscopic mechanism to skew scattering and side jump effects. Unlike the quantum metric dipole mechanism observed in even-layer $MnBi_2Te_4$, the extrinsic spin-orbit coupling in our films introduces spin-flip scattering, an essential component of skew scattering and side jump. By tuning the gate voltage, we observe multiple sign reversals in the nonlinear transverse conductivity, with a strong dependence on the chemical potential. Leveraging this nonlinear response, $MnBi_2Te_4$ devices exhibit a broadband bias-free microwave rectification in the frequency range of 1-8 GHz.

**Results and Discussion**

**Room-temperature nonlinear response in $MnBi_2Te_4$ films**

$MnBi_2Te_4$ is in $R\bar{3}m$ structure with Te-Bi-Te-Mn-Te-Bi-Te septuple layer (SL), as schematically depicted in Fig. 1a. The A-type antiferromagnetic order can be described as intralayer ferromagnetic and interlayer antiferromagnetic coupling. Here, $MnBi_2Te_4$ films with the epitaxial orientation along the c-axis (Supplementary Fig. 1a) were directly grown on $Al_2O_3$ and $SrTiO_3$ substrates via molecular beam epitaxy.[26] To protect $MnBi_2Te_4$ from oxidization, a thin layer of 2-nm Al was grown subsequently. The layered structure with a thickness of ~1.39 nm was confirmed via cross-section transmission electron microscopy (TEM, Fig. 1b). We carried out the anisotropic magnetization measurements of zero-field-cooled (ZFC) and field-cooled (FC) magnetization and field-dependent magnetization along in-plane (H//ab) and out-of-plane (H//c) to explore its magnetic properties. Displayed in Fig. 1c, the Néel temperature ($T_N$) of 20 SL MnBi$_2$Te$_4$ is extracted to be ~23 K with the ground magnetic moment along the c-axis, which is consistent with its A-type antiferromagnetic feature[27,28]. According to the linearly polarized second harmonic generation (SHG) maps of MnBi$_2$Te$_4$, a six-fold pattern is present due to broken surface inversion symmetry, similar to the reported SHG of exfoliated flake[29]. Although defects generated during the growth process may contribute to inversion symmetry breaking, such an effect is much smaller than that of the surface sate-originated SHG. This is consistent with the topological insulator feature of MnBi$_2$Te$_4$ whose surface state has broken inversion symmetry above the Néel temperature[17,18,30]. The intensity of the SHG varies quadratically with laser power, as displayed in Supplementary Note 2 and Fig. 2.

To perform nonlinear transport measurement, MnBi$_2$Te$_4$ films were fabricated into standard Hall-bar devices. The field-dependent longitudinal magnetoresistance and transverse Hall signals of 7 SL MnBi$_2$Te$_4$ at 2K are shown in Fig. 1e, and sheet electron carrier density can be extracted to be $2.74 \times 10^{12}$ cm$^{-2}$. As displayed in Fig. 1f, the anomalous Hall effect (AHE) at various temperatures can be obtained after subtracting the normal Hall part. Because of the A-type AF order with weak interlayer coupling, Mn from different layers can flip their spins independently so the multiple magnetic intermediate spin-flop states with kinks are identified in the AHE curve, which has also been observed in MnBi$_2$Te$_4$ bulk crystals and exfoliated nanoflake.[18,19,31,32] Different from the exfoliated MnBi$_2$Te$_4$ flakes where the even-layer shows compensated antiferromagnetic order with nearly zero Hall signal at zero magnetic field,[19,22,23] both our odd-layer and even-layer thick MnBi$_2$Te$_4$ films exhibit an uncompensated net magnetization with sizeable Hall signal at zero magnetic field below the Néel temperature of ~25 K (Fig. 1f and Supplementary Fig. 3).

Next, we examine the second harmonic signals along transverse and longitudinal directions simultaneously in Hall bar devices in which there was negligible mixing between each channel. All data were collected under zero magnetic field unless specifically mentioned. The device configuration is shown in Fig. 2a. An alternating current $I_{//}$ (ω=13.7 Hz) was applied and the voltage at the second harmonic frequency (ω=27.4 Hz) was measured. We first examined the nonlinear transport in a 7 SL device (Fig. 2b), where the second harmonic transverse voltage $V_{xy}^{2\omega}$ is larger than the longitudinal voltage $V_{xx}^{2\omega}$ by 4 times. Typical curves for $V_{xy}^{2\omega}$ at 10 K and 300 K are shown in Fig. 2c. It is noteworthy that the nonlinear transverse signal can be detected at room temperature. Secondly, $V_{xy}^{2\omega}$ scales quadratically with input current, and it changes sign when the current direction and contact electrodes are inverted. The quadratic relation and sign reversal for the two opposite measurement geometries suggest its origin as a second-order nonlinear response. Figure 2d displays the temperature-dependent nonlinear second harmonic signal. As the longitudinal resistance changes with temperature, we converted the measured signal $V_{xy}^{2\omega}$ and input current $I_{//}$ to electric fields through $E_{//} = \frac{V_{//}}{L_{//}} = \frac{I_{//} \times R_{xx}}{L}$ and $E_{xy}^{2\omega} = \frac{V_{xy}^{2\omega}}{W}$ with L and W of the longitudinal length and transverse width of the device. $E_{xy}^{2\omega}$ scales linearly with $E_{//}^2$. To compare the temperature-dependent nonlinear response quantitatively, the magnitude of the nonlinear transverse response, defined as $\frac{E_{xy}^{2\omega}}{E_{//}^2}$ is used,[9] and plotted in Fig. 2e. $\frac{E_{xy}^{2\omega}}{E_{//}^2}$ first increases as temperature increases, followed by a decreasing trend. It reaches a maximum near the critical Néel temperature of MnBi$_2$Te$_4$, which suggests the nonlinear second-order response can be used for Néel vector detection. Such a phenomenon is similar to the NLHE predicted in antiferromagnet CuMnAs.[24] It can be observed that the longitudinal conductivity $\sigma_{xx}$ and nonlinear second-order response magnitude show similar variation with temperature (Fig. 2f).

**Physical mechanism of high-temperature nonlinear response in MnBi$_2$Te$_4$**

To explore the physical origin of the observed nonlinear second-order response, we study the scaling relation of the magnitude of the nonlinear second-order response $\frac{E_{xy}^{2\omega}}{E_{//}^2}$ with longitudinal conductivity $\sigma_{xx}$ for devices of various thicknesses. As shown in Figs. 3a-b, the nonlinear second-order response shows a linear dependence on the square of conductivity $\sigma_{xx}$ above $T_N$, which is consistent with the scaling equation below

$$\frac{E_{xy}^{2\omega}}{E_{//}^2} = \xi \sigma_{xx}^2 + \eta, \tag{1}$$

where $\xi$ and $\eta$ are constants. Here, the slope $\xi$ stands for the contribution from skew scattering effect, and the intercept $\eta$ quantifies the contribution from side jump and Berry curvature dipole effects.[9,13,33] The threefold rotation symmetry of MnBi$_2$Te$_4$ forbids in-plane Berry curvature dipole.[7] By fitting the data, we obtain $\xi= 6.14\times10^{-18}$ m$^3$V$^{-1}$S$^{-2}$ and $\eta= -1.81\times10^{-10}$ m V$^{-1}$ for 7 SL, which are larger than the 20 SL device ($\xi= 2.27\times10^{-18}$ m$^3$V$^{-1}$S$^{-2}$ and $\eta= -8.76\times10^{-11}$ mV$^{-1}$ for 20 SL). This suggests that both skew scattering and side jump effects are enhanced in thinner devices, which may correlate with the bandgap difference.[34] The opposite sign between $\xi$ and $\eta$ indicates the competing contributions from skew scattering and side jump effects. The deviation of $\frac{E_{xy}^{2\omega}}{E_{//}^2}$ from linear scaling (values collected at around the Néel temperature) in Fig.3a may be due to the magnetic spin fluctuation[35]. To track the symmetry-dependence of the nonlinear signal, angular-dependent measurement in a disc-like device with radially distributed electrodes was further performed. As shown in Supplementary Fig. 6b, the nonlinear second-order response shows a threefold rotational symmetry, fully compliant with the threefold symmetry of MnBi$_2$Te$_4$. Thus, the dominant mechanism is due to skew scattering and side jump effects. The extrinsic effects, like Schottky contact and Joule heating, have been safely excluded which was discussed in detail in Supplementary Note 3. Since the nonlinear conductivity $\sigma_{yxx}^{2\omega}$ can be written as: $\frac{E_{xy}^{2\omega}}{E_{//}^2} = \frac{\sigma_{yxx}^{2\omega}}{\sigma}$, the $\sigma_{yxx}^{2\omega}$ can be obtained through the experimentally determined second order response $\frac{E_{xy}^{2\omega}}{E_{//}^2}$ and conductance $\sigma$. As $\sigma$ is linearly dependent on the scattering time $\tau$, $\sigma_{yxx}^{2\omega}$ will scale with $\tau$ and $\tau^3$.

To gain a deeper understanding of the physical origin of the observed nonlinear transport, we consider the presence of warping in MnBi$_2$Te$_4$. The threefold rotational symmetry detected in the nonlinear transverse signals (angular dependence nonlinear measurements displayed in Supplementary Figs. 6 and 9) suggests a hexagonal warping effect in the band structure[36]. Additionally, an energy splitting develops between the surface states of the same band index but opposite surface momenta upon formation of the long-range magnetic order.[26] The effective Hamiltonian can be written as $H = H_0 + H_E + U(r)$, with the unperturbed Hamiltonian $H_0$ of the band Hamiltonian including hexagonal warping and a gap parameter, $H_E$ of the external driving field and $U(r)$ for the disorder scattering potential. The unperturbed band Hamiltonian $H_0$ has the form

$$H_0 = H_{so} + H_w + H_m = \hbar v_F(k_x \sigma_y - k_y \sigma_x) + i\frac{\lambda}{2}(k_+^3 - k_-^3)\sigma_z + \Delta \sigma_z, \quad (2)$$

where $(k_x, k_y)$ are the wave factors, $(\sigma_x, \sigma_y, \sigma_z)$ are the Pauli matrices, $v_F$ is the Fermi velocity, $\lambda$ is the warping parameter, $\Delta$ is the energy gap of the system, and $k_\pm = k_x \pm ik_y$. Here, $H_{so}$ represents the spin-orbit coupling, $H_w$ stems from the warping effect and $H_m$ represents the gap. The Hamiltonian contains a tunneling (mass) term, assumed to be large, a term accounting for Rashba spin-orbit coupling which is responsible for the Dirac cone of topological insulators, as well as a warping term. The latter is vital in giving rise to a second-order response, since this requires mirror-symmetry breaking. We focus on temperatures above the Néel temperature without considering the magnetism. At these temperatures scattering arises from both phonons and impurities. Both phonons and impurities give rise to momentum scattering as well as spin-dependent scattering, which leads to spin flips. Importantly, in topological materials with strong spin-orbit interaction, the scattering potential itself contains an extrinsic spin-orbit term. Such a term is known to exist in topological insulators, where it is needed to explain the STM quasiparticle interference[37] and affects weak anti-localization[38]. In our case, such an extrinsic spin-orbit term is present in the phonon as well as in the impurity scattering potentials. Determining the correct dependence and prefactors is an involving process requiring a detailed consideration of inter-band dynamics, which is sketched in Supplementary Note 1. Theoretically, we obtain the nonlinear transverse conductivity can be simplified as:

$$\sigma_{yxx}^{2\omega} = \gamma_\beta \left( C_1\, \tau_{\text{imp}} + C_2\, \tau_{\text{imp}}^3 \right) \cos 3\varphi, \tag{3}$$

where $\gamma_\beta = \beta^2 u_0^2 \pi k_F^4 \rho(\varepsilon_F)/\hbar$, $\varphi$ is the angle between the current and x-axis, $\tau_{\text{imp}}$ is the momentum relaxation time. The most important findings are as follows. Firstly, both $C_1$ and $C_2$ are determined by the strength of extrinsic spin-orbit scattering and reflect the presence of both skew scattering and side jump, which are essentially indistinguishable in topological materials[39,40]. Secondly, the joint effect of warping and the presence of a gap makes the system anisotropic[41] for the observation of a nonlinear response. Thirdly, the nonlinear transverse conductivity $\sigma_{yxx}^{2\omega}$ increases with increasing Fermi energy $E_F$, as shown in Fig. 3c. Finally, the temperature dependence is primarily due to phonon scattering, and $\sigma_{yxx}^{2\omega}$ decreases with increasing temperature (Fig. 3d), since $E_F \gg k_B T$ usually for topological materials. The observed nonlinear response is a manifestation of spin-charge locking in TIs. It is well known from the Edelstein[42,43] and spin-Hall effect[44,45] that charge dynamics in systems with strong spin-orbit interactions affect spin dynamics. Here we see the converse effect in which spin-flip dynamics due to SOC influences charge transport. Extrinsic spin-orbit coupling is known to have a strong effect on charge transport, an example being the transition from weak localization to weak anti-localization as a function of extrinsic spin-orbit coupling strength[38]. In our experiments, however, this effect is not related to localization, but rather to the interplay of charge and spin distributions in nonlinear response. The nonlinear second-order response $\sigma_{yxx}^{2\omega}$ as a function of temperature for 7 SL and 20 SL MnBi$_2$Te$_4$ devices are shown in Fig. 3e-f, respectively. It can be fitted with the equation

$$\sigma_{yxx}^{2\omega} = (1 + BT) \left[ \frac{C_1}{(1+AT)} + \frac{C_2}{(1+AT)^3} \right], \tag{4}$$

where coefficient $B$ comes from spin-orbit scattering, $C_1$ consists of side jump and skew scattering contributions, and $C_2$ stems from the skew scattering effect. By fitting the data, we obtain $B= -4.75 \times 10^{-4}$, $C_1= 4.95 \times 10^{-14}$, and $C_2= -4.76 \times 10^{-14}$ for 7 SL ($B= 1.59 \times 10^{-3}$, $C_1= 6.09 \times 10^{-14}$, and $C_2= -6.49 \times 10^{-14}$ for 20 SL). The opposite sign between $\xi$ and $\eta$ is consistent with nonlinear transverse magnitude fitting analysis (Figs. 3a-b).

The comparable value between $C_1$ and $C_2$ for each device suggests that skew scattering and side jump are on the same footing. The factor $(1+BT)$ comes from the fact that phonons also affect the skew scattering mechanism encapsulated in the extrinsic spin-orbit coupling, as shown in Supplemental S1.

According to our theoretical results in Fig. 3c, the nonlinear second-order response varies with chemical potential, thus we further investigated the $V_{xy}^{2\omega}$ as a function of gate voltage ($V_g$) where high-dielectric SrTiO$_3$ was used as a bottom gate. In Fig. 4a, $V_{xy}^{2\omega}$ presents multi oscillations and sign reversals as $V_g$ decreasing from 140 V to -140 V, while the longitudinal resistance $R_{xx}$ increases monotonously (Fig. 4b). The tunability of $V_{xy}^{2\omega}$ decreases with temperature (Fig. 4a), consistent with the lower $R_{xx}$ tunability at high temperatures (Fig. 4b), which is due to the decreased dielectric constant of SrTiO$_3$. Two typical Hall curves at $V_g= \pm 140$ V are shown in Fig. 4c inset, whereby the negative slope suggests the Fermi energy level lies in the conduction band throughout the whole measurement. From the change of $R_{xx}$, we replot the $V_g$-dependent $V_{xy}^{2\omega}$ in the form of nonlinear transverse conductivity $\sigma_{yxx}^{2\omega}$, as displayed in Fig. 4c. The multiple sign-reversals of $\sigma_{yxx}^{2\omega}$ with chemical potential can be clearly observed. This phenomenon also helps to exclude the thermoelectric effect that only changes sign with the carrier type. Electron-dominated conduction is preserved in the whole $V_g$ range of 160 V to -160 V with the electron density changing from $1.99 \times 10^{13}$ to $5.50 \times 10^{12}$ cm$^{-2}$ (Fig. 4d).

**Room-temperature RF rectification in MnBi$_2$Te$_4$**

The second-harmonic nonlinear response has the potential to be utilized for high-frequency rectification purposes. To test MnBi$_2$Te$_4$ device as a rectifier, we exploit its ability to perform wireless microwave rectification at room temperature, without requiring any external bias or magnetic field. As illustrated in Fig. 5a, the device is subjected to an electromagnetic wave emitted by a patch antenna. Employing the same measurement setup used for the nonlinear transport characterization, we measure the rectified DC voltage in both the transverse ($V_{xy}^{dc}$) and longitudinal ($V_{xx}^{dc}$) directions.

MnBi$_2$Te$_4$ demonstrates rectification capabilities across 1 to 8 GHz, with the best performance observed within the 1-2 GHz range, as depicted in Fig. 5b. To further illustrate the device's response, Figure 5c presents a 2D mapping of $V_{xy}^{dc}$ as a function of RF frequency and power, demonstrating its broad bandwidth response. Figure 5d indicates that the rectification voltage $V_{xy}^{dc}$ exhibits a linear increase with the power of the RF source. When the measurement configuration is switched to the longitudinal direction, the measured $V_{xx}^{dc}$ also observed a linear relation with the RF power, but its magnitude is ~ 2.5 times smaller than $V_{xy}^{dc}$, as shown in Figs. 5e-f. Since the material can be scaled up in area via MBE growth, further improvement in rectification performance can be achieved through appropriate antenna design and optimized impedance matching circuits.

In summary, we have demonstrated a room-temperature nonlinear transverse response in topological uncompensated antiferromagnet MnBi$_2$Te$_4$ films that is independent of Berry curvature or quantum metric dipole. The quantum metric dipole previously reported for the antiferromagnet phase of MnBi$_2$Te$_4$ operates below Neel temperature[22,23] and does not contribute to the second-order nonlinear response observed here. Instead, scaling analysis revealed that skew scattering and side jump effects associated with the anisotropic warping effect introduce the nonlinear transverse response that is robust even at room temperature. MnBi$_2$Te$_4$ can rectify broadband RF radiation over the range of 1-8 GHz without the application of external bias. Our study highlights the general contribution of warping to second-order nonlinear transport, which manifests strongly at room temperature when other intrinsic mechanisms vanish.

**Methods**

**MnBi$_2$Te$_4$ film growth and device fabrication**

MnBi$_2$Te$_4$ films on (0001) Al$_2$O$_3$ and (111) SrTiO$_3$ substrates were prepared by molecular beam epitaxy through the co-evaporating three fluxes of Mn, Bi, and Te.[26] Before taking samples out of the chamber, 2 nm Al was grown as the capping layer. The confined Hall-bar devices were fabricated by the standard electron beam lithography and reactive-ion etching processes, and Pt contacts were deposited via e-beam evaporation.

**Electrical and RF rectification measurement**

The nonlinear transport measurement was performed in a Lake Shore Cryostat (CRX-VF probe station), with a Keithley 6221 source meter used to apply alternating current to $MnBi_2Te_4$ device and lock-in amplifiers of SR830 to collect the voltage signals. $SrTiO_3$ substrate served as a bottom gate with the gate voltage applied by Keithley 2636B. The magneto-transport experiment was carried out in an Oxford TeslatronPT system equipped with a multi-channel lock-in amplifier. The phases of measured first- and second-harmonic voltage are confined to ~0° and 90°, respectively. For the RF rectification measurement, an SG384 upgraded with an output frequency doubler was used to generate the RF signals ranging from 1 GHz to 8 GHz, and an antenna and Keithley 2182A nanovoltmeter were used to transmit the microwave to $MnBi_2Te_4$ devices and collect the rectified DC signals. The magnetization measurements were performed in DC-Superconducting-Quantum-Interface-Devices.

**Data availability**

Relevant data supporting the key findings of this study are available within the article and the Supplementary Information file. The raw data that support the plots within this study are available from the corresponding author upon request.

**Acknowledgments**

K.P.L. acknowledges the support from Singapore's National Research Foundation, Prime Minister's Office, Singapore under Competitive Research Program Award NRF-CRP22-2019-0006. F.X. was supported by the National Natural Science Foundation of China (52350001, 52225207, and 11934005), the Shanghai Pilot Program for Basic Research - FuDan University 21TQ1400100 (21TQ006), and the Shanghai Municipal Science and Technology Major Project (Grant No.2019SHZDZX01).

**Author contributions**

K.P.L. conceived and supervised the project. S.L. fabricated the devices and performed the transport measurement. Q.Z. and J.C. performed the magnetization measurement. C.L., R.Z., Q.X., and K.L. performed the SHG measurement. N.W. and S.L. measured the RF rectification under the supervision of W.G.. S.L., X.-B.Q., Z.Z.D, and K.P.L. analyzed the data. R.B. and D.C. performed the theoretical calculations. E.Z. and F.X. grew the MnBi$_2$Te$_4$ films. S.L., R.B., D.C., and K.P.L. wrote the manuscript, with input from all authors.

**Ethics declarations**

Competing interests
The authors declare no competing interests.

**Figure captions**

**Fig. 1 | Atomic structure and basic characterization of schematic picture of MnBi₂Te₄. a,** Crystal structure from side view (top) and top view (bottom). **b,** Cross-section high-resolution transmission electron microscopy. The layer thickness is extracted to be ~1.39 nm. Scale bar: 1 nm. **c,** Zero-field-cooled (ZFC) and field-cooled (FC) magnetization along c-axis and ab-plane. The Néel temperature ($T_N$) is estimated to be ~23 K. The applied magnetic field is 5000 Oe. Inset, Magnetization hysteresis along the c-axis and ab-plane measured at 1.8 K. **d,** Polar-plot of second harmonic generation (SHG) data at 300 K, represented by black dots, confirming the broken surface inversion symmetry. The red dash lines are fitting to the equation in J.F. et.al[29]. **e,** Longitudinal magnetoresistance ($R_{xx}$) and Hall data ($R_{xy}$) under out-of-plane magnetic field measured at 2 K. The inset is the measurement configuration with in-plane current $I$ and out-of-plane magnetic field $\mu_0 H$. **f,** Anomalous Hall data ($R_{AHE}$) at different temperatures. Multiple spin-flop steps originate from the A-type antiferromagnetism.

**Fig. 2 | Observation of room-temperature nonlinear second-order response in 7 septuple layer (SL) MnBi₂Te₄. a,** Schematic geometry of the nonlinear transport measurement. Gold, blue, and brown elements indicate the Pt contact, MnBi₂Te₄ sample, and Al₂O₃ substrate, respectively. **b**, Second harmonic nonlinear response along transverse ($V_{xy}^{2\omega}$) and longitudinal direction ($V_{xx}^{2\omega}$). The inset indicates the measurement configuration. **c,** $V_{xy}^{2\omega}$ as a function of current. Dash lines are the quadratic fittings. $V_{xy}^{2\omega}$ depends quadratically on the current and changes sign when the current direction reverses. The insets indicate the different measurement configurations. **d,** Typical temperature-dependent $E_{xy}^{2\omega}$ versus $E_{//}^2$. $E_{xy}^{2\omega}$ shows a linear dependence on $E_{//}^2$ in the measured temperature range of 10 K-310 K. **e,** The magnitude of nonlinear transverse response $\frac{E_{xy}^{2\omega}}{E_{//}^2}$ as a function of temperature. The error bars are added based on the standard fitting error of $E_{xy}^{2\omega}$ versus $E_{//}^2$. **f,** Temperature-dependent longitudinal conductivity.

**Fig. 3 | Scaling of the nonlinear second-order response with longitudinal conductivity.** Magnitude of nonlinear transverse response $\frac{E_{xy}^{2\omega}}{E_{//}^2}$ as a function of the square of conductivity $\sigma_{xx}^2$ for **a,** 7 SL and **b,** 20 SL MnBi₂Te₄ devices. The error bars are added based on the standard fitting error of $E_{xy}^{2\omega}$ versus $E_{//}^2$. **c,** The calculated

nonlinear transverse conductivity ($\sigma_{yxx}^{2\omega}$) as a function of Fermi energy ($E_F$) for bandgap $\Delta$=15 meV, 20 meV and 25 meV at zero temperature. We take the Fermi velocity $v_F$=1.6×10$^5$ m/s, the warping constant $\lambda$ = 80 eVÅ$^3$, impurity relaxation time $\tau_{imp}$=0.12 ps and scattering time $1/\gamma_\beta$ =0.12 ps. **d,** The calculated $\sigma_{yxx}^{2\omega}$ as a function of temperature for impurity relaxation time $\tau_{imp}$=0.12 ps, 0.125 ps and 0.13 ps. Here, Fermi energy, bandgap, and scattering time $1/\gamma_\beta$ are fixed to be 25 meV, 15 meV and 1.2 ps, respectively. **e, f,** Temperature-dependent $\sigma_{yxx}^{2\omega}$ for 7 SL and 20 SL devices. The dash lines are the theoretical fitting results. The error bars are added based on the standard fitting error of $E_{xy}^{2\omega}$ versus $E_{//}^2$.

**Fig. 4 | Gate-tunable nonlinear transport. a**, Second harmonic nonlinear response along transverse ($V_{xy}^{2\omega}$) a function of gate voltage ($V_g$). **b,** The longitudinal resistance ($R_{xx}$) as a function of $V_g$. Inset, typical Hall data under $V_g$ = ±140 V. The sample shows electron-dominated conduction property in the whole applied $V_g$. **c,** The corresponding nonlinear transverse conductivity $\sigma_{yxx}^{2\omega}$ as a function of $V_g$, with multiple sign-reversal. This helps to exclude the thermoelectric effect. **d,** Electron carrier density (n) and mobility (μ) as a function of $V_g$ measured at 7.6 K. The error bars are added based on the standard fitting error of Hall data at different $V_g$.

**Fig. 5 | Room-temperature wireless rectification. a**, Schematic illustration of rectification measurement on MnBi$_2$Te$_4$ device. **b,** The rectified DC voltage along the transverse direction ($V_{xy}^{dc}$). It shows a broad response among 1-8 GHz. **c,** 2D mapping plot of rectification signal as a function of power ($P_{IN}$) and frequency ($f_{IN}$). The highest $V_{xy}^{dc}$ corresponds to the frequency of 1.7 GHz. **d,** $V_{xy}^{dc}$ as a function of the radiofrequency power. **e,** The rectified DC voltage along the longitudinal direction ($V_{xx}^{dc}$). **f,** $V_{xx}^{dc}$ versus radio frequency power.

**Table 1 The second-harmonic nonlinear signal in different materials**

| Materials | Input current maximum (µA) | Output voltage maximum (µV) | Highest working temperature (K) | References |
|---|---|---|---|---|
| Bilayer WTe$_2$ | 1 | 200 | 100 | Ref.[46] |
| Few-layer WTe$_2$ | 600 | 30 | 100 | Ref.[9] |
| Strained monolayer WSe$_2$ | 4.5 | 20 | 140 | Ref.[47] |
| Twisted bilayer WSe$_2$ | 0.04 | 20,000 | 30 | Ref.[11] |
| Corrugated bilayer graphene | 0.1 | 2 | 15 | Ref.[10] |
| Bi$_2$Se$_3$ | 1,500 | 20 | 200 | Ref.[13] |
| Bulk WTe$_2$ | 110 | 4,000 | 4.2 | Ref.[48] |
| Cd$_3$As$_2$ | 110 | 4,000 | 4.2 | Ref.[48] |
| Ce$_3$Bi$_4$Pd$_3$ | 10,000 | 0.8 | 4 | Ref.[49] |
| T$_d$-MoTe$_2$ (c-axis) | 5,000 | 40 | 40 | Ref.[50] |
| T$_d$-MoTe$_2$ (in-plane) | 97 | 125 | 100 | Ref.[51] |
| TaIrTe$_4$ | 600 | 120 | 300 | Ref.[14] |
| α-(BEDT-TTF)$_2$I$_3$ | 1,000 | 9 | 40 | Ref.[52] |
| Pt | 50 | 120 | 375 | Ref.[16] |
| hBN/graphene/hBN moiré superlattice | 5 | 100 | 220 | Ref.[53] |
| 4SL MnBi$_2$Te$_4$ flake | 10 | 200 | 20 | Ref.[23] |
| 6SL MnBi$_2$Te$_4$/BP heterostructure | 10 | 350 | 20 | Ref.[22] |
| BaMnSb$_2$ | 100 | 250 | 400 | Ref.[15] |
| BiTeBr | 5 | 100 | >350 | Ref.[12] |
| MnBi$_2$Te$_4$ films | 250 | 225 | >320 | This work |

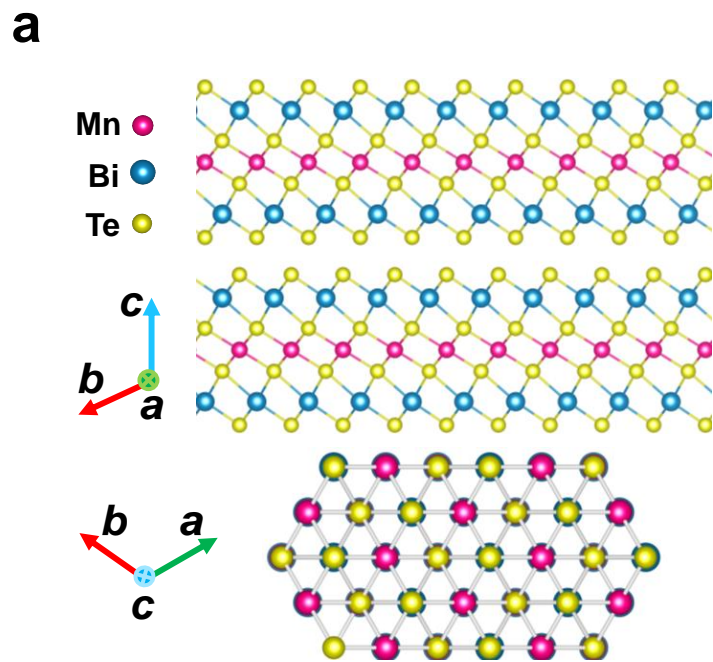
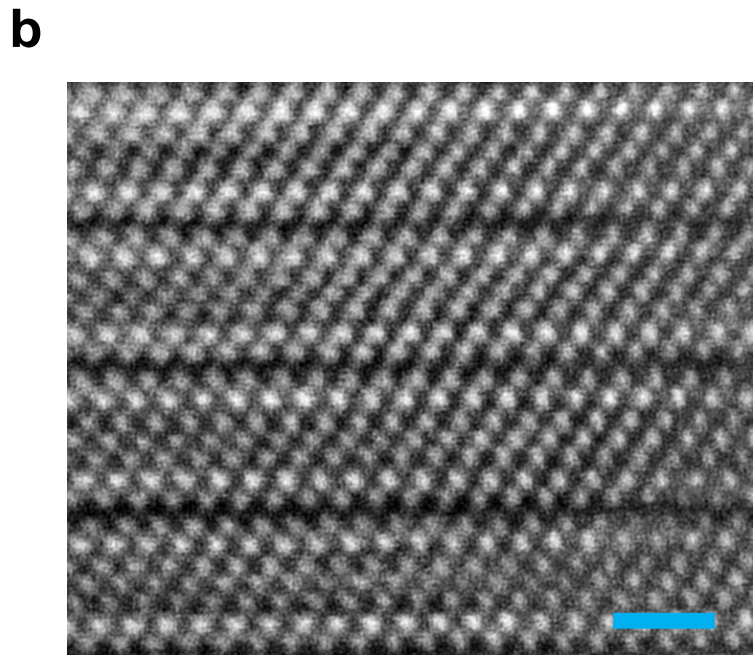
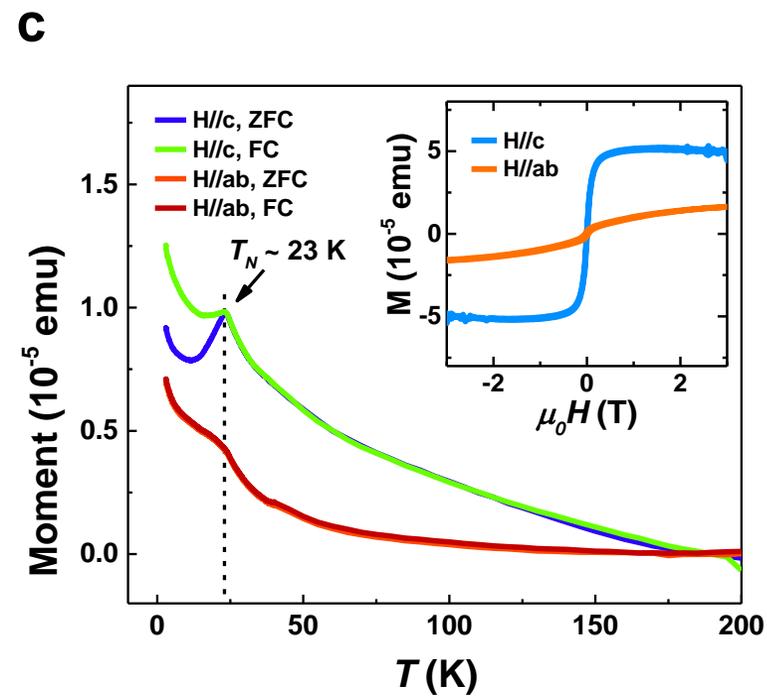
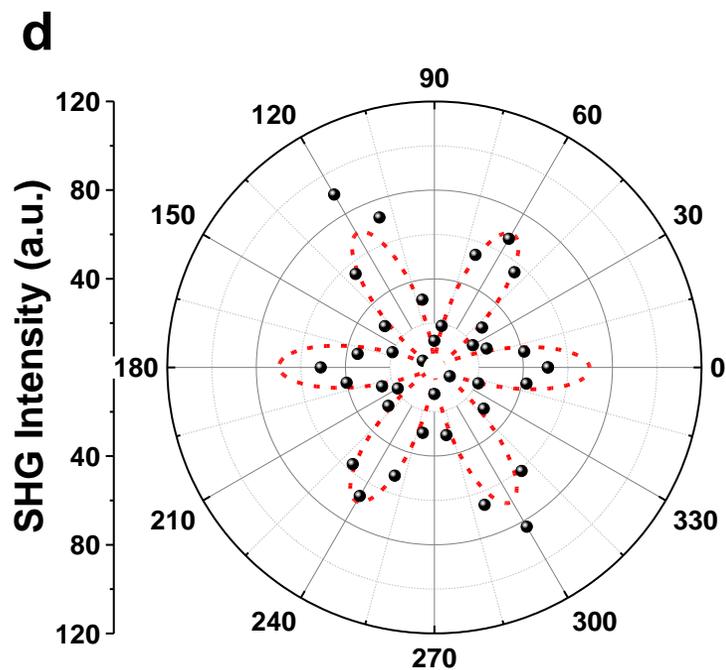
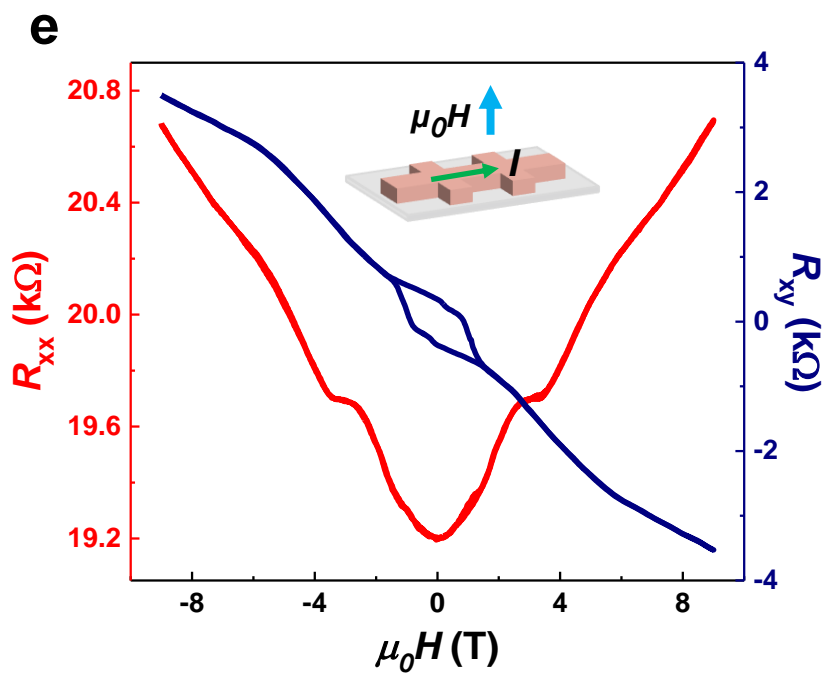
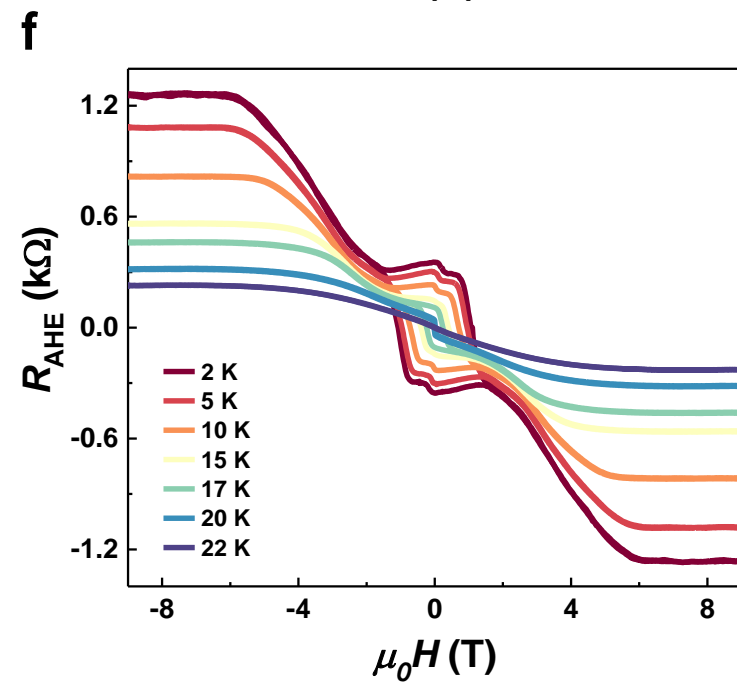

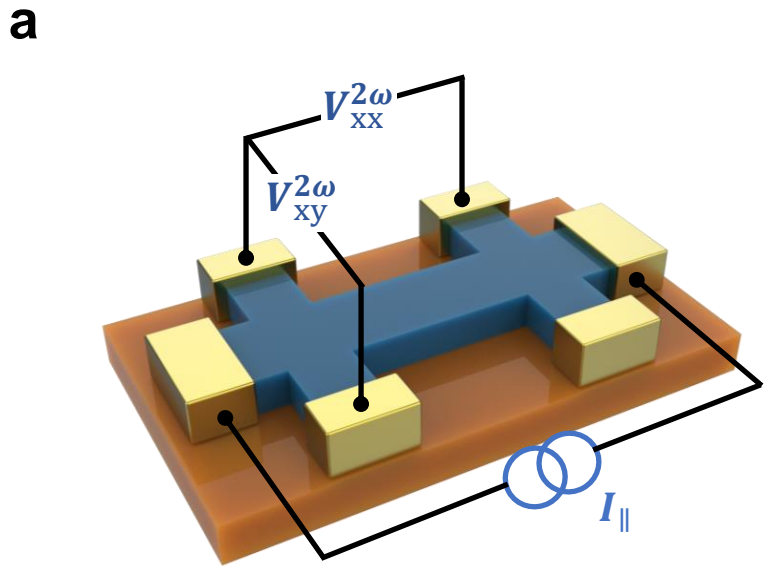
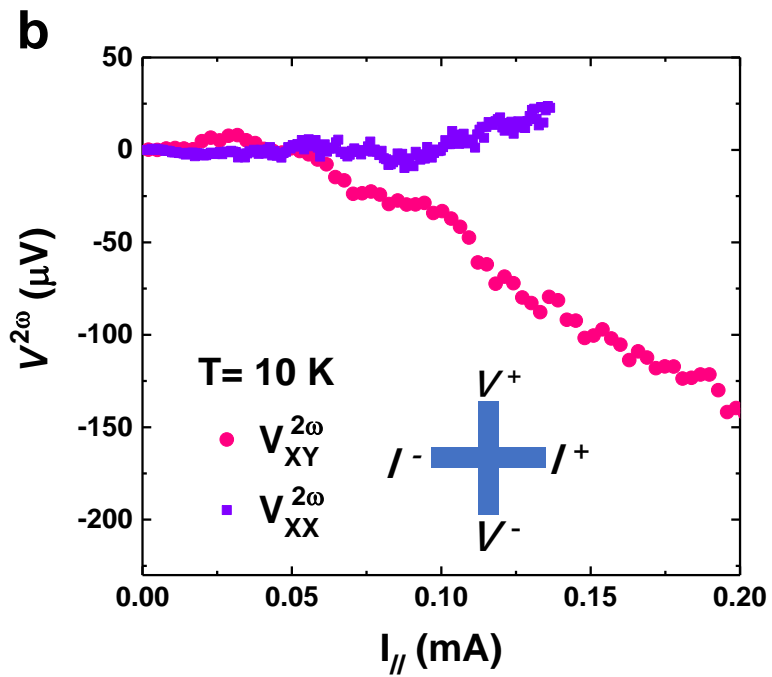
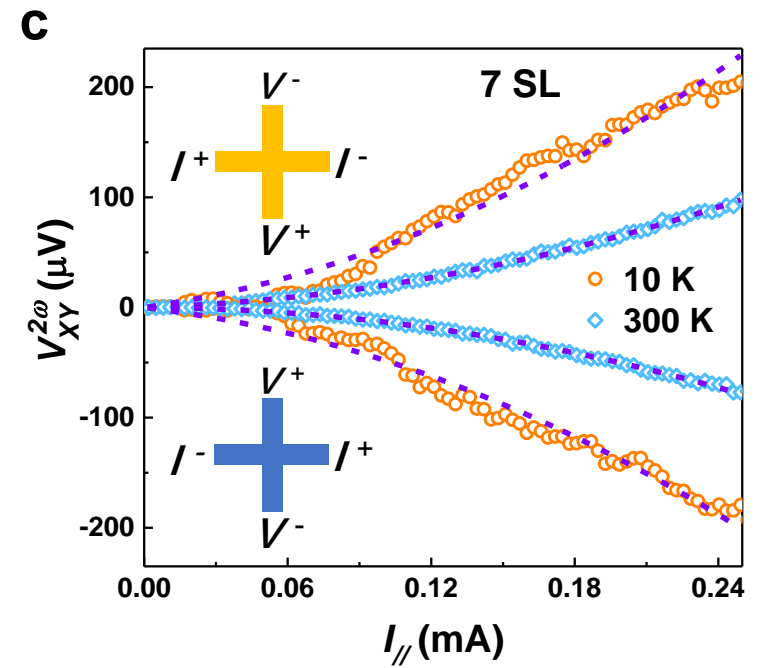
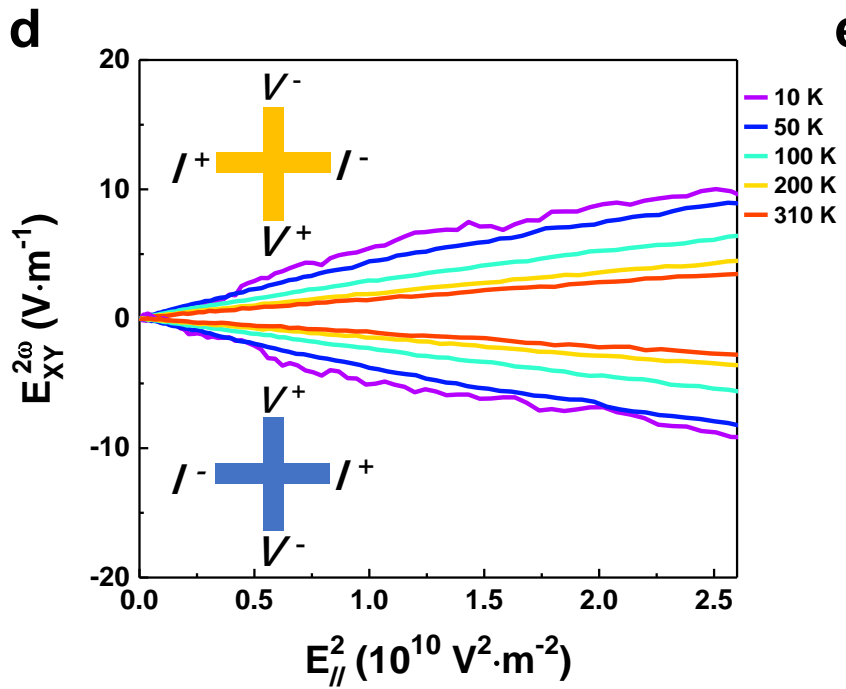
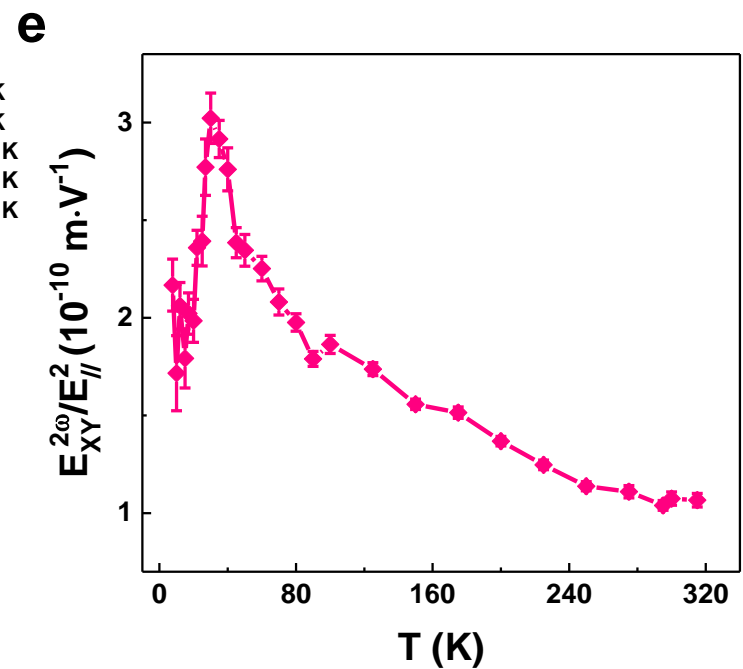
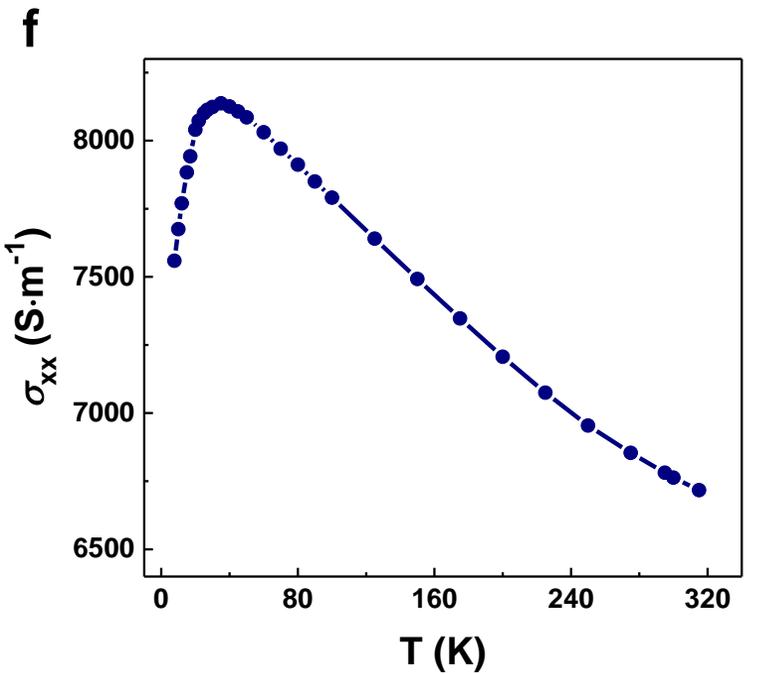

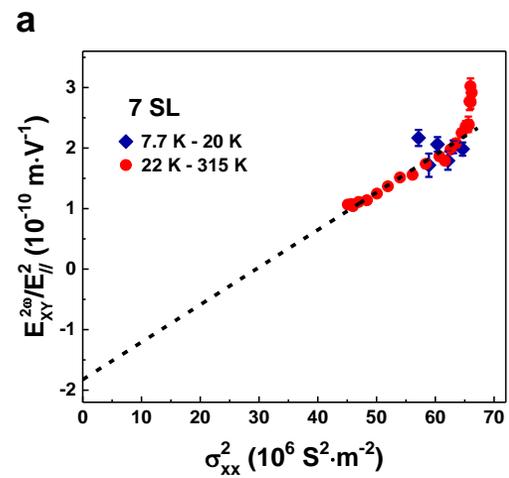
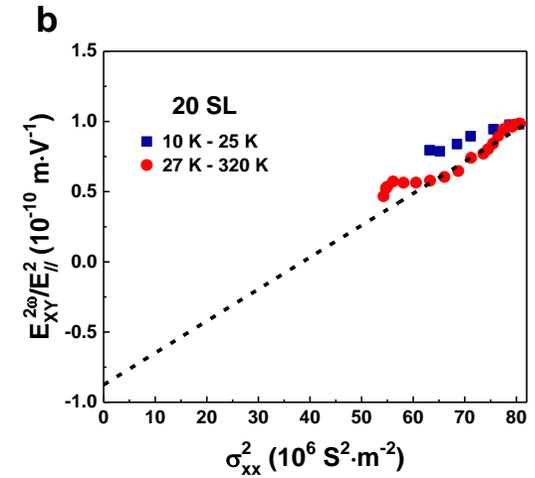
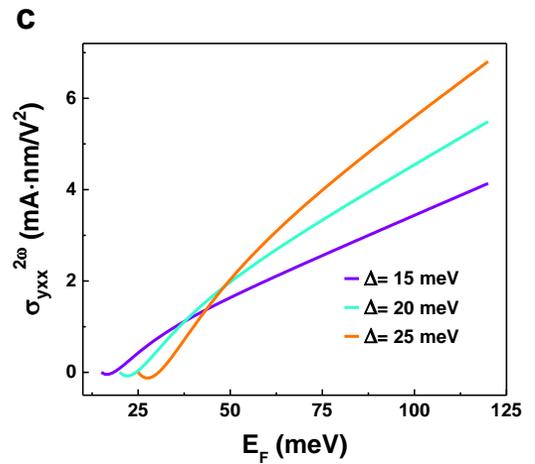
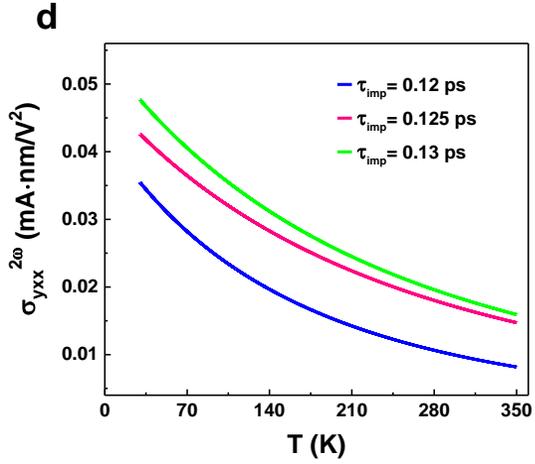
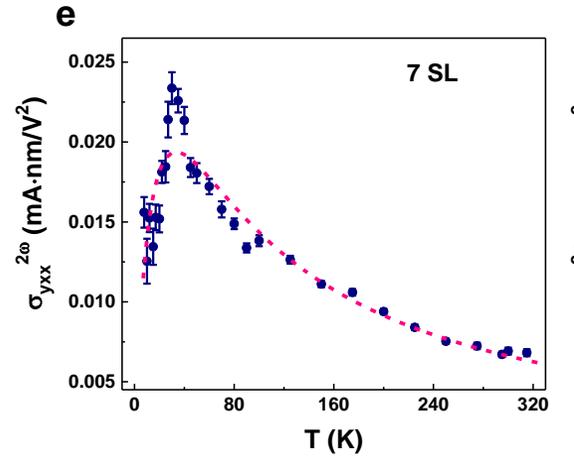
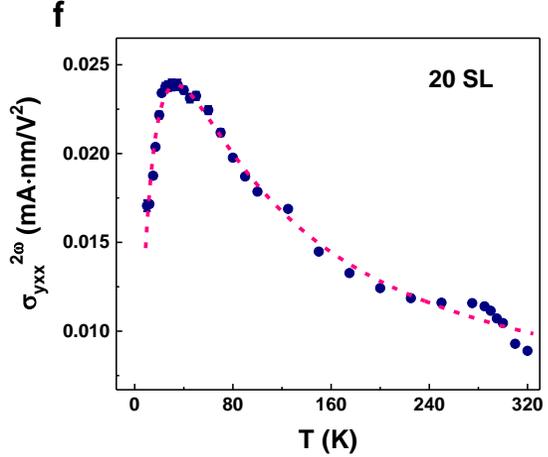

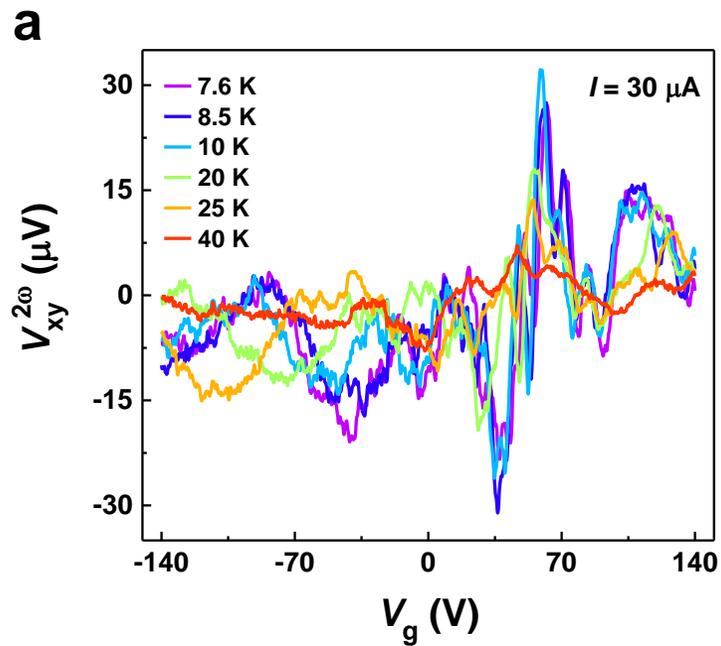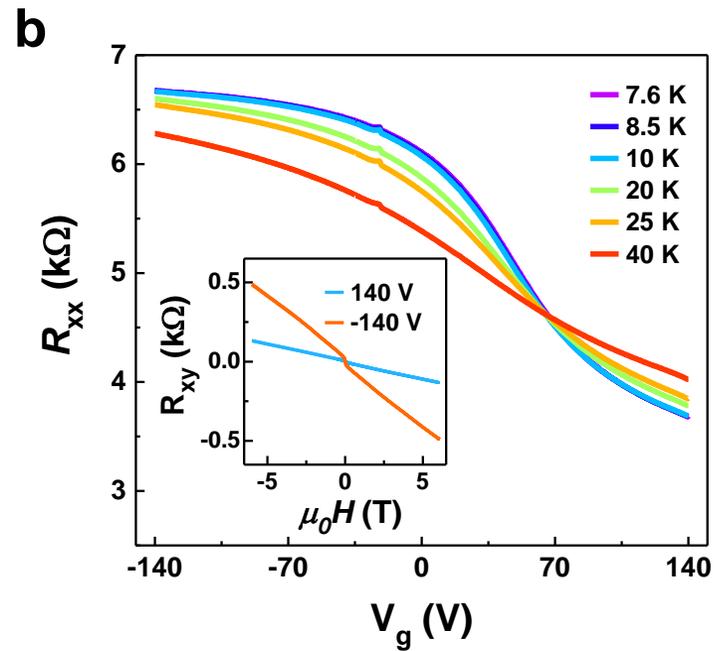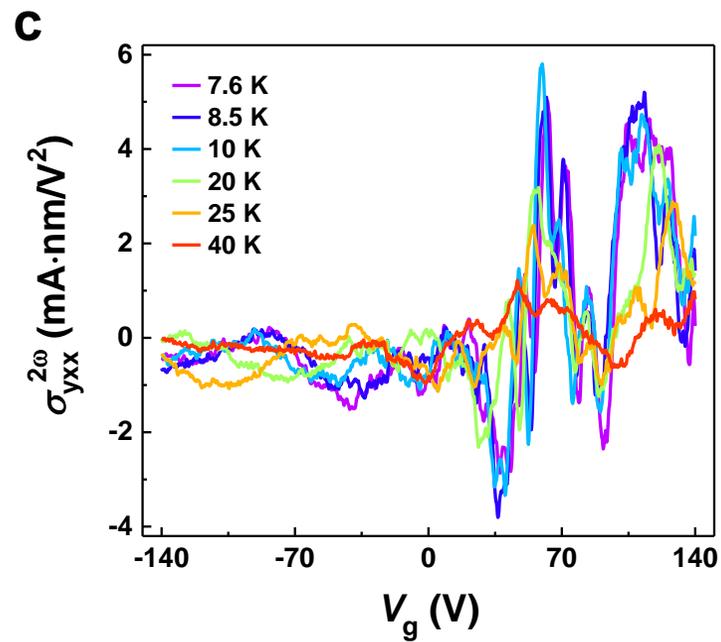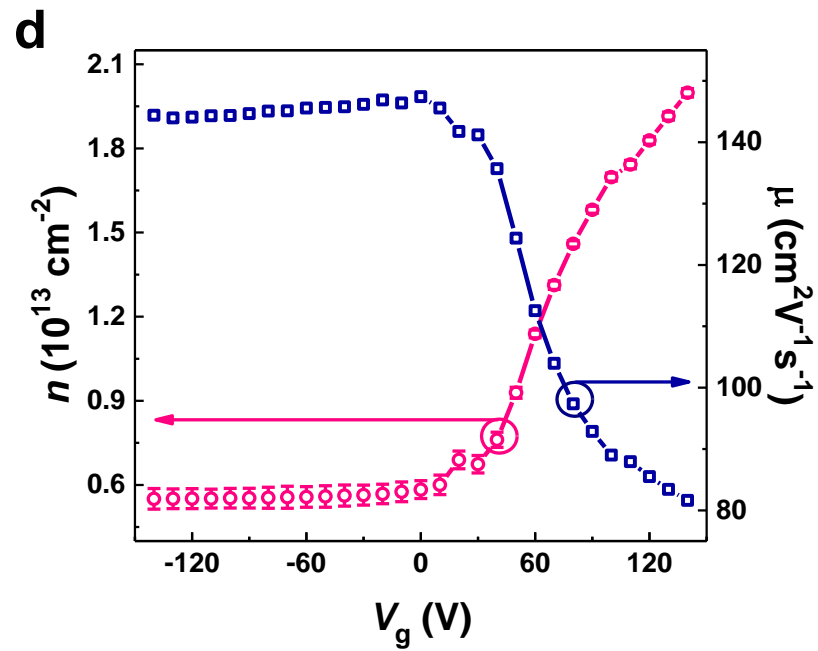

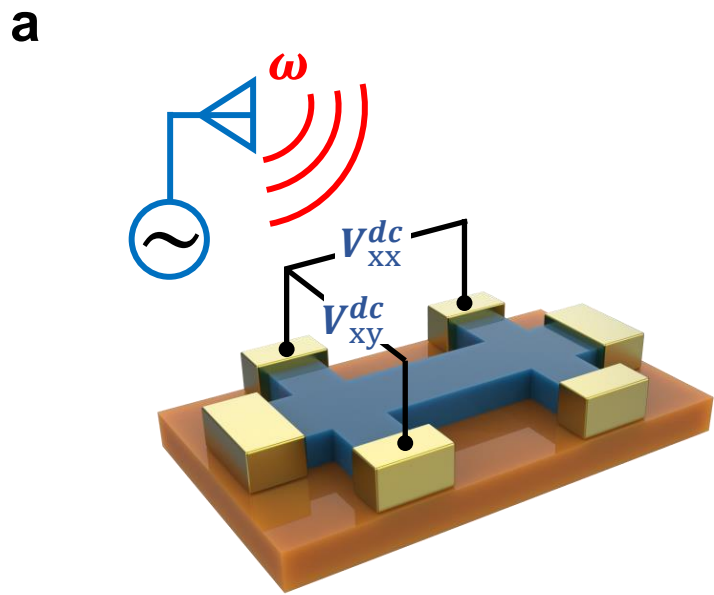
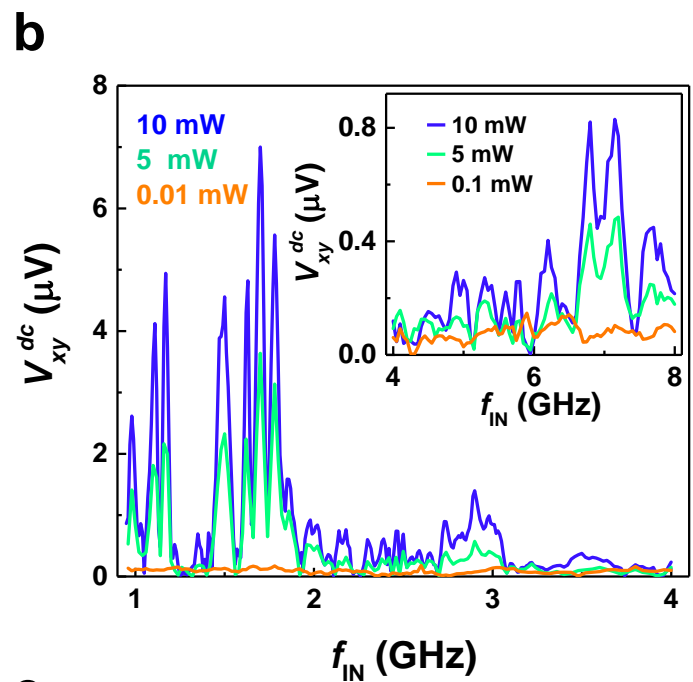
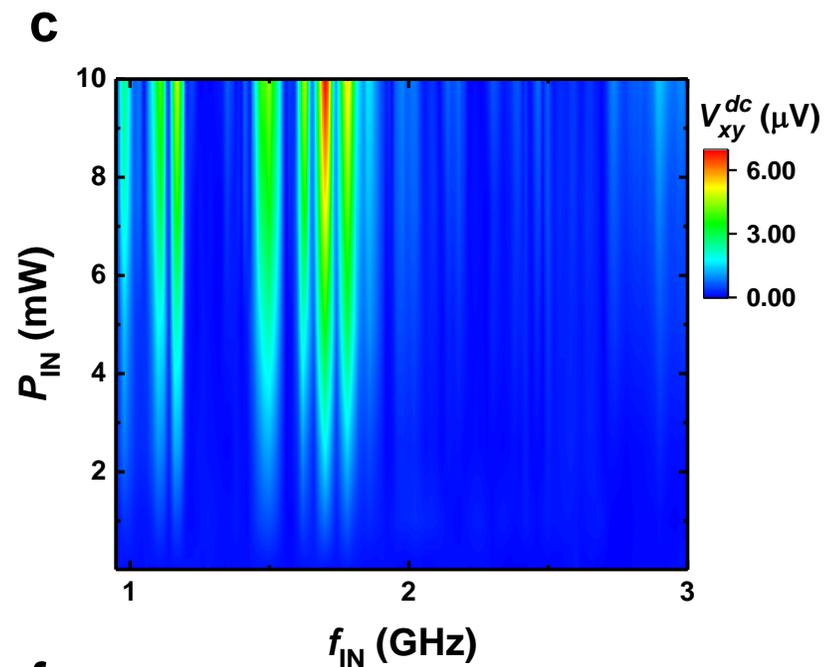
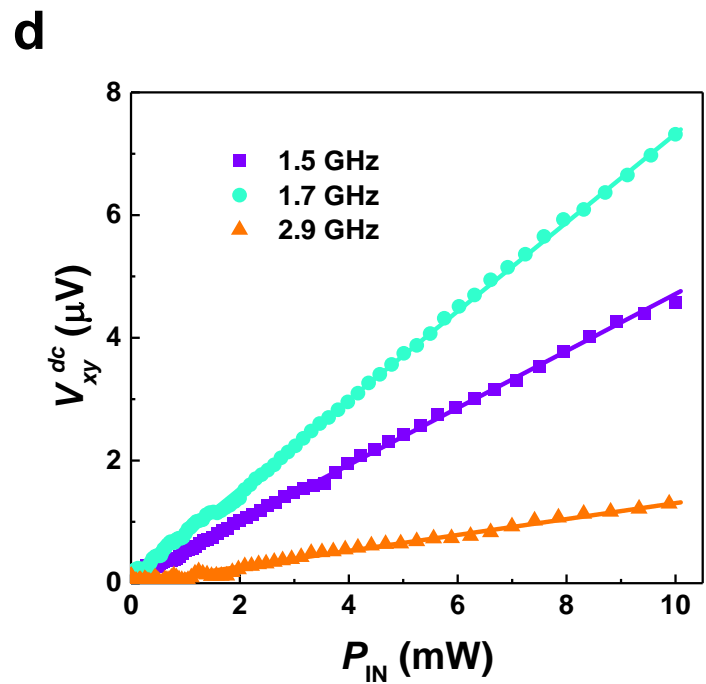
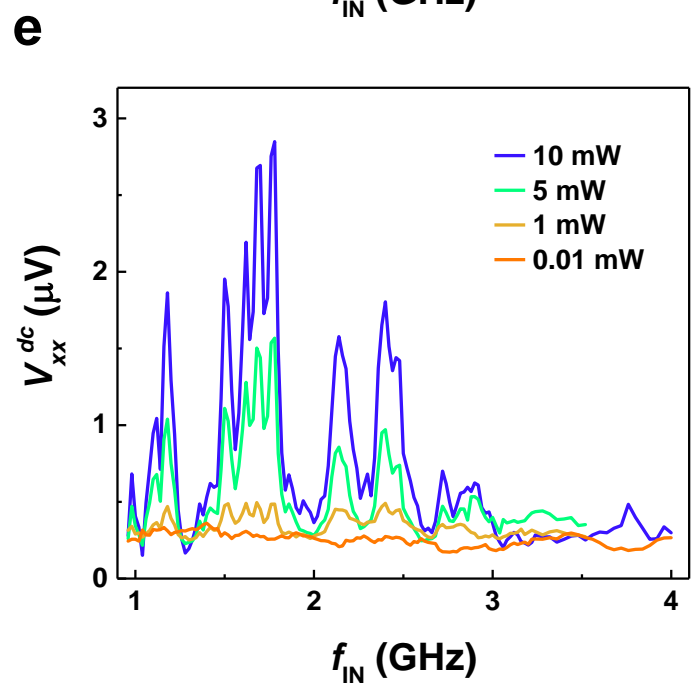
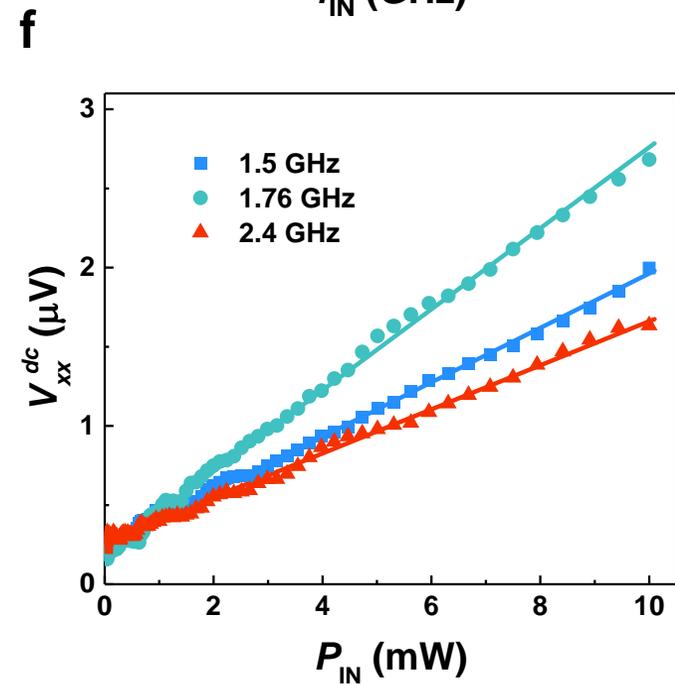